\documentclass[runningheads]{llncs}
\usepackage{graphicx}
\usepackage{booktabs}
\usepackage{multirow}
\usepackage{amssymb}
\usepackage{amsmath}
\usepackage{makecell}
\usepackage[colorlinks]{hyperref}
\usepackage{color}

\begin{document}
\title{Modality-aware Mutual Learning for Multi-modal Medical Image Segmentation}
%
%
\author{Yao Zhang\inst{1, 2} \and Jiawei Yang\inst{3} \and Jiang Tian\inst{4} \and Zhongchao Shi\inst{4} \and Cheng Zhong\inst{4} \and Yang Zhang\inst{5,\dag} \and Zhiqiang He\inst{1,5,\dag}}

\authorrunning{Yao Zhang, Jiawei Yang et al.}
%
\institute{Institute of Computing Technology, Chinese Academy of Sciences, Beijing, China \and University of Chinese Academy of Sciences, Beijing, China \and Electrical and Computer Engineering, University of California, Los Angeles, USA \and AI Lab, Lenovo Research, Beijing, China \and Lenovo Corporate Research \& Development, Lenovo Ltd., Beijing, China}
\renewcommand{\thefootnote}{}
\maketitle              

\footnote{$\dag$: Equal contribution as the corresponding authors. This work is done when Yao Zhang was an intern at AI Lab, Lenovo Research.}

\begin{abstract}
Liver cancer is one of the most common cancers worldwide. Due to inconspicuous texture changes of liver tumor, contrast-enhanced computed tomography (CT) imaging is effective for the diagnosis of liver cancer. In this paper, we focus on improving automated liver tumor segmentation by integrating multi-modal CT images. To this end, we propose a novel mutual learning (\textbf{ML}) strategy for effective and robust multi-modal liver tumor segmentation. Different from existing multi-modal methods that fuse information from different modalities by a single model, with ML, an ensemble of modality-specific models learn collaboratively and teach each other to distill both the characteristics and the commonality between high-level representations of different modalities. The proposed ML not only enables the superiority for multi-modal learning but can also handle missing modalities by transferring knowledge from existing modalities to missing ones. Additionally, we present a modality-aware (\textbf{MA}) module, where the modality-specific models are interconnected and calibrated with attention weights for adaptive information exchange. The proposed modality-aware mutual learning (\textbf{MAML}) method achieves promising results for liver tumor segmentation on a large-scale clinical dataset. Moreover, we show the efficacy and robustness of MAML for handling missing modalities on both the liver tumor and public brain tumor (BRATS 2018) datasets. Our code is available at \href{https://github.com/YaoZhang93/MAML}{https://github.com/YaoZhang93/MAML}.

\end{abstract}
\section{Introduction}
Liver cancer is one of the most common cancer diseases in the world~\cite{bray2018global}. CT images are the most commonly used imaging modality for the initial evaluation of liver cancer. The accurate measurements of liver tumor status from CT images, including tumor volume, shape, and location, can assist doctors in making hepatocellular carcinoma evaluation and surgical planning. However, a portion of textures of the liver tumor on CT volumes are inconspicuous and, therefore, can be easily neglected even by experienced radiologists. In clinical practice, radiologists usually enhance CT images by an injection protocol for clearly observing liver tumors. When the contrast agent goes through the liver within blood vessels, it yields a favorable contrast between liver tissues and abnormalities, including liver tumors. Contrast-enhanced CT imaging used in the dual-modality protocol is comprised of venous and arterial phases with intravenous contrast delay.
Dual-phase images can make good complementary information for each other and thus can contribute to better diagnosis of liver tumor.

In recent years, deep learning has largely advanced the field of computer-aided diagnosis (CAD), especially medical image segmentation~\cite{ronneberger2015u,milletari2016v,zhu2019anatomynet}. Fully Convolution Neural Networks (FCNs) go beyond the limitation of hand-crafted features and dramatically improve the performance of liver tumor segmentation with an encoder-decoder architecture~\cite{liu20183d,zhang2019light,li2018h,zhang2021darn,zhang2018sequentialsegnet}. There exists two major issues applying FCNs in multi-modal segmentation. One is how to integrate information from multi-modal medical images effectively. The other is how to deal with the scenario of missing modalities that is common in practice. We elaborate them in the followings.

Multi-modal information has been fused and applied for different purposes, e.g., brain segmentation~\cite{zhou2019hyper}, diagnosis~\cite{liang2020oralcam}, and 3D dental reconstruction~\cite{liang2020x2teeth,song2021oral}, which is also extended to CT images. Most methods extend the single-modal method to a multi-stream model, where each stream is intended for a specific modality. The modality-specific features extracted by different streams are fused in subsequent modules. Notably, the input multi-modal images should be registered before feeding into the model. Based on the encoder-decoder architecture, the strategies for multi-modal feature fusion can be classified into four categories. 
The first one is an early-fusion strategy, where multi-modal images are integrated at the input and processed jointly along a single stream of network~\cite{isensee2021nnu}. Second, instead of merging both phases at the input of the network, a middle-fusion strategy processes different modalities independently in the corresponding encoders, and these modalities share the same decoder for feature fusion and final segmentation~\cite{chen2019octopusnet}. Third, a late-fusion fashion makes each phase go through an independent stream of an encoder-decoder network, and the learned features are fused at the end of each stream~\cite{sun2017automatic}. At last, an ultimate one introduces hyper-connections between and within encoder-decoder networks to enable more effective information exchange between different modalities~\cite{zhou2019hyper}. However, in these methods, the features from each modality are straightforwardly combined, and consequently, the diverse contribution of different modalities is neglected. 

The strategies proposed to handle missing modalities include synthesizing missing modalities by a generative model~\cite{orbes-arteaga2018simultaneous} or learn a modality-invariant feature space~\cite{dorent2019hetero,havaei2016hemis}. However, synthesizing missing modalities requires heavy computations, and existing modality-invariant methods usually failed when most of the modalities are missing. Recent KD-Net~\cite{hu2020knowledge} transfers knowledge from a multi-modal network from a mono-modal one by knowledge distillation. However, KD-Net relies on one student model for each missing modality and an additional teacher model to perform only one-way knowledge transfer to the student model, which brings extra computation cost and limits the multi-modal representation.

\begin{figure}[!tp]
 \centering
 \includegraphics[width=11cm, height=5.5cm]{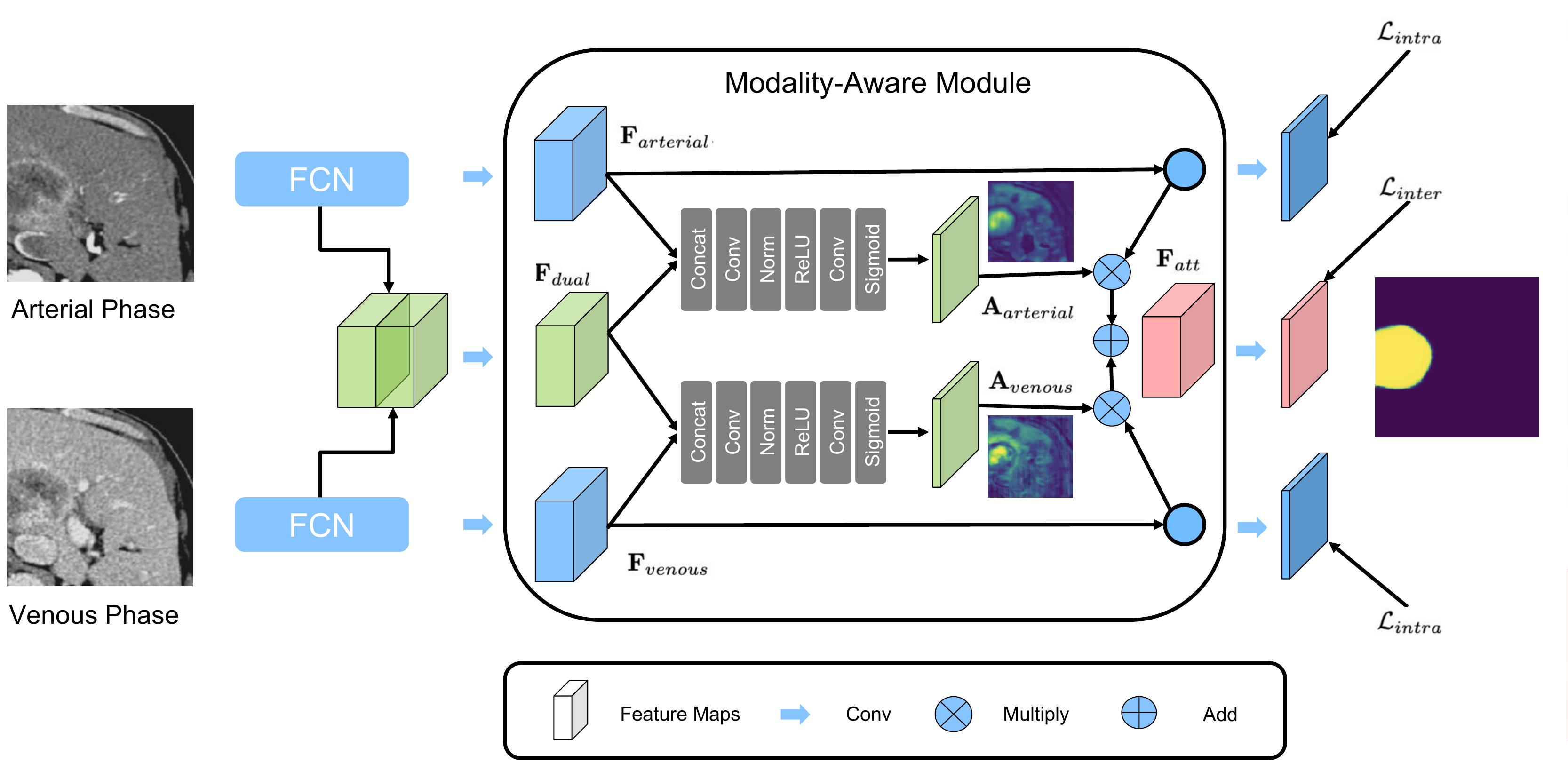}	
 \caption{Illustration of the framework. The input multi-modal CT images are first embedded by different modality-specific FCNs. Then a modality-aware module regresses attention maps, which reflect how to achieve an effective and interpretable fusion of the modality-specific features. The framework is trained by mutual learning strategy composed of intra- and join losses.}
 \label{attention}
 \end{figure}

In this paper, we present a novel Modality-aware Mutual Learning (\textbf{MAML}) method for effective and robust multi-modal liver tumor segmentation.
Specifically, we construct a set of modality-specific models to handle multi-modal data, where each model is intended for one modality. To enable more effective and interpretable information exchange across different modalities, we carefully design an Modality-aware (\textbf{MA}) module to adaptively aggregate the model-specific features in a learnable way. For each model, MA module produces weight maps to value the features pixel by pixel, and then the features are merged by a weighted aggregation for effective multi-modal segmentation.
Moreover, inspired by~\cite{zhang2018deep} and ~\cite{hu2020knowledge}, we design an novel Mutual Learning (\textbf{ML}) strategy. Different from~\cite{hu2020knowledge}, ML enables interactive knowledge transfer to improve the generalization ability of a model and avoid the use of superfluous teacher model. With ML, modality-specific models solve the task collaboratively. We achieve this by training the models through intra-modality and joint losses: the former encourages each model to learn discriminative modality-specific features, while the latter encourages each model to learn from each other to keep the commonality between high-level features for better incorporation of multi-modal information. To sufficiently leverage the deep learning method's power, we collect a large-scale clinical dataset with 654 CT volumes to evaluate the proposed method. Experimental results demonstrate that the proposed MAML significantly outperforms other advanced multi-modal works by a remarkable margin. 
Specifically, MAML reports a promising performance of 81.25\% in terms of Dice per case for liver tumor segmentation. 
Moreover, on the clinical dataset and public BRATS 2018 dataset, we show the effectiveness and robustness of MAML for handling missing modalities in an extreme scenario where only one modality is available.
 
\section{Method}
\label{m}
MAML employs a set of modality-specific models to collaboratively and adaptively incorporates both arterial and venous phase images for accurate liver tumor segmentation. In this case, it consists of two modality-specific models to learn specific features in each modality and a MA module to explore correlated features between two modalities adaptively. Note that the proposed method can be easily extended for more modalities.

\subsection{Modality-specific Model}
A modality-specific model is a common FCN for single-modal segmentation. As UNet~\cite{ronneberger2015u} has been proven successful in medical image segmentation, MAML adopts the powerful nnUNet model~\cite{isensee2021nnu}, one of the state-of-the-art UNet-like framework for medical image segmentation, to achieve the feature extraction from raw CT images. 
The input of dual-phase CT volumes individually goes through each model, and the high-level semantic embeddings of specific phases from the last layer are obtained. It is worth to note that the high-level semantic embeddings share the same shape of the input image. The outputs of different modality-specific models are denoted as $\mathbf{F}_i \in \mathbb{R}^{C \times D \times H \times W}$, where $C = 32$ is the number of channels, $D,H,W$ are the depth, height, and width, and $i \in \{AP,VP\}$. $AP$ and $VP$ are the abbreviations for arterial and venous phases respectively. 
 
\subsection{Modality-Aware Module}
\label{pam}

As illustrated in Fig.~\ref{attention}, we propose an MA module via an attention mechanism to adaptively measure the contribution of each phase. The attention model is widely used for various tasks, including semantic segmentation~\cite{fu2019dual}. Several attention mechanisms have been proposed to enhance the representation of network~\cite{zhang2019et,wang2019automatic,schlemper2019attention,tian2019automatic,chen2017amc}. In this study, we explore the cross-modality attention mechanism to selectively highlight the target features embedded in a single modality to obtain more discriminative dual-modal features for liver tumor segmentation.
 
The outputs of modality-specific models are concatenated together along channels to generate $\mathbf{F}_{dual}$ by a followed convolution layer. Although $\mathbf{F}_{dual}$ encodes both arterial and venous information of liver tumor, it also inevitably introduces redundant noise from each modality for liver tumor segmentation. Instead of obtaining straightforward segmentation from $\mathbf{F}_{dual}$, we propose MA via attention mechanism to adaptively measure each phase's contribution and visually interpret it. 
 
MA module leverages $\mathbf{F}_{dual}$ and $\mathbf{F}_i$ as inputs and produces $\mathbf{F}_{att}$. Specifically, we first generate an attention map $\mathbf{A}_i$ for each $\mathbf{F}_i$, which indicates the significance of the features in $\mathbf{F}_{dual}$ for each specific phase. Given the $\mathbf{F}_i$ of each phase, we concatenate them with the $\mathbf{F}_{dual}$, and then produce the attention weights $\mathbf{A}_i$:
\begin{equation}
	\begin{aligned}
	\mathbf{A}_i = \sigma(f_a([\mathbf{F}_{dual}; \mathbf{F}_i];\theta_i)), i \in \{AP,VP\},
	\end{aligned}
\end{equation}
where $\sigma$ is a Sigmoid function, and $\theta$ represents the parameters learned by $f_a$, which consists of two cascaded convolutional layers. The first convolutional layer uses $3 \times 3 \times 3$ kernels, and the second convolutional layer applies $1 \times 1 \times 1$ kernels. Each convolutional layer is followed by an instance normalization~\cite{ulyanov2016instance}, and a leaky rectified linear unit (Leaky ReLU). These convolutional operations are employed to model the correlation of the discriminative dual-modality information with respect to the features of each modality.

Then, we multiply the attention map $\mathbf{A}_i$ with the $\mathbf{F}_i$ in an element-wise manner. $\mathbf{F}_{att}$ is calculated by a weighted sum of each $\mathbf{F}_i$, defined as:
\begin{equation}
	\begin{aligned}
	\mathbf{F}_{att} = \sum_{i \in \{AP, VP\}} \mathbf{A}_i * \mathbf{F}_i.
	\end{aligned}
\end{equation}

We apply the MA module for each phase to selectively emphasize their characteristics. During this process, the attention mechanism is used to generate a set of attention maps to indicate how much attention should be paid to the $\mathbf{F}_i$ for more discriminative $\mathbf{F}_{att}$. Furthermore, those attention maps provide a visual interpretation of the contribution of each phase for liver tumor segmentation, which is crucial in clinical practice. 
 
\subsection{Mutual Learning Strategy}
\label{mtst}

The learning of the set of modality-specific models is formulated as a voxel-wise binary classification error minimization problem with respect to the ground-truth mask. We carefully design the ML strategy for multi-modal liver tumor segmentation. Concretely, each modality-specific model interacts as a teacher and a student mutually. Thus, the venous model not only draws clues for tumor segmentation from the venous phase but also learns from the arterial model and vice versa. To achieve this, we introduce an intra-phase loss and a joint one. The former encourages each stream to learn discriminative phase-specific features, while the latter encourages each stream to learn from each other to keep the commonality between high-level features for better incorporation of multi-modal information. Let $X = \{X_{venous}, X_{arterial}\}$ be the input venous and arterial volumes respectively, $Y$ be the ground-truth annotations, and $W = \{W_{venous}, W_{arterial}\}$ be the weights in venous and arterial streams respectively. The goal of the teacher-student training scheme is to minimize the following objective function
\begin{equation}
	\begin{aligned}
		\mathcal{L} = \lambda\sum_{i \in \{AP, VP\}} \mathcal{L}_{intra}(Y|X_i; W_i) + \mathcal{L}_{joint}(Y|X; W),
	\end{aligned}
\end{equation}
where both intra-phase loss $\mathcal{L}_{intra}$ and joint loss $\mathcal{L}_{joint}$ are standard segmentation loss function, and $\lambda$ is the weight factors that are empirically set as 0.5. We employ a combination of Cross-Entropy loss and Dice loss as the segmentation loss to reduce the effect of imbalanced data distribution of tumors.

The advantages of ML lie in the following three aspects: (1) it enables the model to be capable of dealing with both multi-modal segmentation and handling missing modalities without any modification, which is applicable and efficient in clinical practice; (2) each model for single modality can implicitly leverage dual-modality information by learning from the other models, which leads to better segmentation results even when other modalities are missing; (3) combined with characteristics and commonality of each modality, the collaboration of all model-specific models can make a better multi-modal segmentation.

\section{Experiments and Results}
\label{er}
\begin{table*}[!t]
\centering
\caption{Results on multi-modal liver tumor segmentation. Best results are highlighted with bold.}
\label{tab:multimodal}
\setlength{\tabcolsep}{9mm}{
\begin{tabular}{l|c|c}
\toprule
Methods  & Dice {[}\%{]} $\uparrow$ & ASSD {[}voxel{]} $\downarrow$ \\ \midrule \midrule
nnUNet~\cite{isensee2021nnu}                  & 78.76 $\pm$ 18.91       & 8.02 $\pm$ 20.21 \\ \hline 
OctopusNet~\cite{chen2019octopusnet}                 & 78.89 $\pm$ 18.65       & 12.67 $\pm$ 42.43         \\ \hline 
MS+Ensemble                   & 78.96 $\pm$ 19.37       & 5.88 $\pm$ 10.73         \\ \hline
MS+MA                   & 80.98 $\pm$ 18.58       & 5.38 $\pm$ 9.20                   \\ \hline
MAML             & \textbf{81.25 $\pm$ 17.02}       & \textbf{4.71 $\pm$ 6.13}                       \\ \bottomrule
\end{tabular}%
}
\end{table*}
\begin{table*}[!t]
\centering
\caption{Results on handling missing modalities for liver tumor segmentation. Best results are highlighted with bold.}
\label{tab:singlemodal}
\setlength{\tabcolsep}{6mm}{
\begin{tabular}{c|l|c|c}
\toprule
\multicolumn{2}{c|}{Methods} & Dice {[}\%{]} $\uparrow$ & ASSD {[}voxel{]} $\downarrow$\\ \midrule\midrule
\multirow{2}{*}{\makecell[c]{Arterial \\ Phase}} & nnUNet~\cite{isensee2021nnu}                & 71.21 $\pm$ 25.87       & 9.51 $\pm$ 28.34         \\ \cline{2-4} 
                              & MAML                  & \textbf{79.55 $\pm$ 19.06 }        & \textbf{6.38 $\pm$ 12.00}         \\ \hline
\multirow{2}{*}{\makecell[c]{Venous \\ Phase}} & nnUNet~\cite{isensee2021nnu}                  & 75.10 $\pm$ 20.65       & 9.26 $\pm$ 30.82         \\ \cline{2-4}
& MAML      & \textbf{79.81 $\pm$ 18.42}      & \textbf{6.35 $\pm$ 12.03}          \\ \bottomrule
\end{tabular}%
}
\end{table*}
\begin{figure}[!tp]
 \centering
 \includegraphics[width=\linewidth, height=4.2cm]{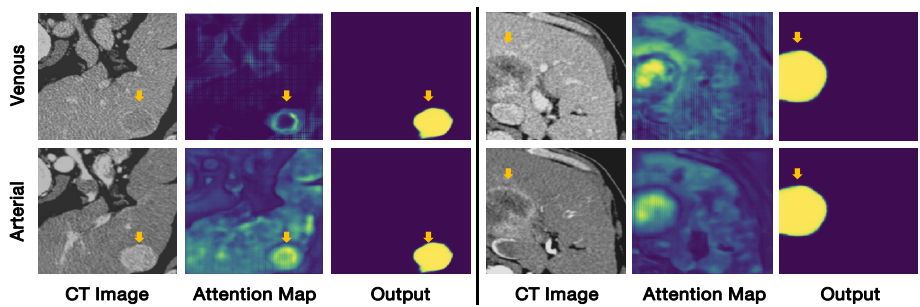}	
 \caption{Attention maps produced by Modality-Aware Module are able to capture enhanced part (left) as well as bleeding part and pseudo capsule (right) of the tumor.}
 \label{attentionmap}
\end{figure}
\textbf{Datasets and Evaluation Metrics.} Experiments are conducted on contrast-enhanced CT volumes obtained from Chinese PLA General Hospital. We acquire 654 contrast-enhanced CT volume\footnote{One volume corresponds to one phase from a patient} with arterial and venous phases. All CT volumes are obtained using SIEMENS scanners. The in-plane size of CT volumes is $512 \times 512$ with spacing ranges from $0.56$ mm to $0.91$ mm, and the number of slices ranges from $67$ to $198$ with spacing $1.5$ mm. Three experienced clinicians from hepatobiliary surgery with extensive experience interpreting the CT images have been involved for validation. To effectively combine multi-modal CT images, we utilize a registration method~\cite{klein2010elastix} to obtain the spatial relation between the images of different phases. For data pre-processing, we truncate the raw intensity values within the range $0.5\%$-$99.5\%$ of the initial HU value and normalize each raw CT case to have zero mean and unit variance. 
BraTS 2018 dataset~\cite{menze2015the} contains MR scans from $285$ patients with four modalities: T$1$, T$2$, T$1$ contrasted-enhanced (T$1$ce) and Flair. The goal of the dataset is to segment three sub-regions of brain tumors: whole tumor (WT), tumor core (TC), and enhancing tumor (ET).
The metrics employed to quantitatively evaluate segmentation include Dice Similarity Coefficient (Dice) and Average Symmetric Surface Distance (ASSD).
\begin{figure}[!tp]
 \centering
 \includegraphics[width=\linewidth, height=3.5cm]{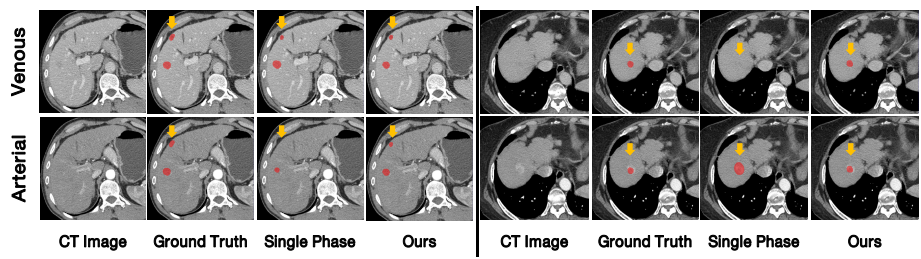}	
 \caption{Qualitative examples where our method detects the tumors while single-phase methods fail on arterial phase (left) or venous phase (right). The tumors are marked in red and highlighted with yellow arrows.}
 \label{visualization}
\end{figure}

\noindent\textbf{Implementation Details.} The framework is built with PyTorch on an Nvidia Tesla V100 GPU. The network is trained by the Adam optimizer with an initial learning rate of $0.0003$. Due to the constraint of GPU memory, each volume is sliced into patches with a size of $128 \times 128 \times 128$ before feeding into the network. The network is trained for $600$ epochs, about $150$ hours. 
No further post-processing strategies are applied as we only intend to evaluate the effectiveness of the network design. 
For data augmentation, we adopt on-the-fly random mirroring and rotation, deformation, and gamma correction for all training data to alleviate the over-fitting problem.

\begin{table*}[!t]
\centering
\caption{Results on handling missing modalities for brain tumor segmentation. The results of HeMIS, U-HVED, and KD-Net are derived from~\cite{hu2020knowledge}, where the standard deviations of HeMIS's and U-HVED's results are not provided. Dice is employed for evaluation.}
\label{tab:missing}
\setlength{\tabcolsep}{4.5mm}{
\begin{tabular}{l|c|c|c}
\toprule
Methods & Enhanced Tumor & Tumor Core & Whole Tumor\\ \midrule \midrule
HeMIS~\cite{havaei2016hemis}                & 60.8        & 58.5     & 58.5     \\ \hline 
U-HVED~\cite{dorent2019hetero}                  & 65.5        & 66.7      & 62.4    \\ \hline
KD-Net~\cite{hu2020knowledge}                  & 71.67 $\pm$ 1.22       & 81.45 $\pm$ 1.25       &  76.98 $\pm$ 1.54      \\ \hline
MAML                  & \textbf{73.42 $\pm$ 1.10}      & \textbf{83.36 $\pm$ 1.23}     & \textbf{78.32 $\pm$ 1.41}         \\ \bottomrule
\end{tabular}%
}
\end{table*}

\noindent\textbf{Effectiveness of Multi-modal Modeling.} To demonstrate the effectiveness of MAML, we make an ablation study for MA and ML respectively on the clinical dataset, where one fifth images is for testing, and the rest are for training. The baseline is a straightforward average of the outputs of modality-specific models, denoted as ``MS+Ensemble''. Then we apply MA to aggregate the modality-specific models adaptively, denoted as ``MS+MA''. Finally, we combine both MA and ML, denoted as ``MAML''. As shown in Table~\ref{tab:multimodal}, MA outperforms the baseline in terms of both Dice and ASSD. Moreover, ML further boosts the performance with a remarkable margin. The experimental results demonstrate the effectiveness of MAML for multi-modal liver tumor segmentation. Then we compare MAML with recent advanced methods for multi-modal segmentation, nnUNet~\cite{isensee2021nnu} and OctopusNet~\cite{chen2019octopusnet}. The former takes a concatenation of both phases as input while the latter individually encodes each phase and generate segmentation by one decoder. The results in Table~\ref{tab:multimodal} revealing the outstanding performance of MAML.

\noindent\textbf{Interpretable Fusion.} MA offers not only an effective fusion of different modalities, but also an interpretable one. We illustrate the interpretability by qualitatively visualizing the learned attention map. From Fig.~\ref{attentionmap} (left), we can observe that the venous attention map focuses on the edge of the tumor while the arterial attention map focuses on the body. Besides, a certain number of the tumors' surface and the adjacent liver is usually delineated with a pseudo capsule. In Fig.~\ref{attentionmap} (right), the venous attention map focuses on the pseudo capsule and the bleeding part inside the tumor. It proves that MA can capture the knowledge of medical imaging for an interpretable multi-modal liver tumor segmentation.

\noindent\textbf{Handling Missing Modalities.}
A superiority of ML strategy is the capability of dealing with missing modalities in multi-modal segmentation. We consider an extreme scenario that only one modality is available. On the clinical dataset, the CT images with either arterial or venous phase are available at inference procedure. We set nnUNet, the counterpart of the modality-specific model in MAML, as a baseline and train it solely on arterial or venous phase. From Table~\ref{tab:singlemodal}, it is observed that MAML significantly outperforms the baseline. Besides, the performance gap between arterial and venous phases of MAML is significantly smaller than that of nnUNet, revealing the excellent ability of ML that transfers knowledge between modalities. We also compare MAML with methods specialized for dealing with missing modalities. Following~\cite{hu2020knowledge}, a $3$-fold cross-validation on public BRATS 2018 dataset using only the T1ce modality as input. The results of KD-Net~\cite{hu2020knowledge}, U-HVED~\cite{dorent2019hetero}, and HeMIS~\cite{havaei2016hemis}, in terms of Dice, are directly taken from~\cite{hu2020knowledge}. From Table~\ref{tab:missing}, we observe that our method excels in the other three advanced methods, demonstrating the effectiveness of MAML for handling missing modalities. 
The limitation of the proposed framework in the current implementaiton is that it allows either for the full set of modalities or only one modality as input. We would like to enhance it for arbitrary number of missing modalities in the future work. 

\section{Conclusion}
\label{c}
In this study, we propose MAML that enables effective and robust multi-modal segmentation. ML achieves an ensemble of modality-specific models collaboratively learning the complementary information. MA performs in an adaptive and explainable way for better multi-modal liver tumor segmentation. We illustrate that MAML can substantially improve the performance of multi-modal segmentation and effectively handle missing modalities, which is of great value in clinical practice.

\bibliographystyle{splncs04}
\bibliography{paper1888.bib}
\end{document}